# Additively manufactured Ni-20Cr to V functionally graded material: computational predictions and experimental verification of phase formations

Beril Tonyali[a,1], Hui Sun[a,1], Brandon Bocklund[b], John Paul Borgonia[c], Richard A. Otis[c], Shun-Li Shang[a], Zi-Kui Liu[a], Allison M. Beese [a,d,*]

[a] *Department of Materials Science and Engineering, Pennsylvania State University, University Park, PA 16802, United States*

[b] *Lawrence Livermore National Laboratory*, Livermore, CA 94550, United States

[c] *Engineering and Science Directorate, Jet Propulsion Laboratory, California Institute of Technology, Pasadena, CA 91109, United States*

[d] *Department of Mechanical Engineering, Pennsylvania State University, University Park, PA 16802, United States*

[1] *Both authors have equally contributed to the present work.*

* *Corresponding author: beese@matse.psu.edu*

**Abstract**

A database for the Cr-Ni-V system was constructed by modeling the binary Cr-V and ternary Cr-Ni-V systems using the CALPHAD approach aided by density functional theory (DFT)-based first-principles calculations and *ab initio* molecular dynamics (AIMD) simulations. To validate this new database, a functionally graded material (FGM) using Ni-20Cr and elemental V was fabricated using directed energy deposition additive manufacturing (DED AM) and experimentally characterized. The deposited Ni-20Cr was pure fcc phase, while increasing the amount of V across the gradient resulted in the formation of sigma phase, followed by the bcc phase. The experimentally measured phase data was compared with computational predictions made using a Cr-Ni-V thermodynamic database from the literature as well as the database developed in the present work. The newly developed database was shown to better predict the experimentally observed phases due to its accurate modeling of binary systems within the database and the ternary liquid phase, which is critical for accurate Scheil calculations.

**Keywords:** metals and alloys; intermetallics; laser processing; rapid-solidification; molecular dynamics simulations; thermodynamic modeling

# 1. Introduction

Laser-based directed energy deposition (DED) additive manufacturing (AM) is a process in which three-dimensional components are fabricated layer-by-layer. In powder-based DED, powder feedstock is deposited into a melt pool created by a laser, where the powder melts, cools, and solidifies, fusing to the material below. DED can be used to fabricate functionally graded materials (FGMs), in which composition is tailored as a function of position within a build [1]. The local compositions and thermal histories dictate the phases formed, which in turn determine the spatially varying properties [2,3]. Thus, FGMs provide a means for creating multifunctional components [4] or smooth transitions between compositions of two dissimilar materials to avoid the deleterious effects of abrupt interfaces [5]. However, mixing alloys in molten form during DED can promote the formation of brittle intermetallic phases during solidification or subsequent thermal cycling, which can lead to stresses that result in cracking [1].

Joining iron alloys with titanium alloys has been of interest in the aerospace and nuclear industries where multiple materials must be used within components to achieve required functionality. For example, stainless steel and titanium alloys are both used in aircraft engines where joining them allow for the assembly of engine blades into a singular component [4,6]. Similarly, these same materials are used in the nuclear industry to join components for processing and transporting spent nuclear fuel [7]. However, the direct joining of stainless steels and Ti-6Al-4V by welding, brazing, riveting, diffusion bonding, or AM result in weak joints between the materials due to abrupt changes in elastic or thermal properties and the formation of intermetallics [6], often leading to cracking. Alternatively, to eliminate abrupt interfaces, researchers have studied gradually grading between these terminal alloys using DED AM but have shown that when directly grading between Ti-6Al-4V and stainless steel, brittle intermetallic Fe-Ti and Ni-Ti phases form along the gradient [8–10] and residual stresses accumulate [11], both of which result in cracking of the FGM.

To avoid the formation of intermetallics in FGMs, intermediate elements or alloys may be introduced along the composition pathway between terminal alloys. In a recent study by the present team [12], a gradient scheme that used V, Cr, and Ni-20Cr as intermediate compositions between Ti-6Al-4V and stainless steel was designed, and the CALculation of PHAse Diagrams (CALPHAD) method was used to predict deleterious phases under both equilibrium and rapid solidification conditions. The gradient pathway was fabricated, avoiding intermetallics, and the phases formed along the deposited compositions were found to match those predicted computationally. However, the intermediate gradient path of V to Cr to Ni-20Cr graded only through binary element compositions, which were accurately represented in the thermodynamic database by Choi et al. [13,14]. Also, although free of deleterious phases, cracking was observed in this FGM in the Cr-rich compositions, which was found to have high crack susceptibility, as proven through the crack susceptibility criteria [15]. Additionally, it is undesirable to grade into pure elements due to their poor mechanical performance [15] in comparison to alloys. To eliminate the need to grade into pure elements, grading through ternary Cr-Ni-V compositions could be an alternative composition pathway. However, there is limited experimental and computational modeling on these intermediate compositions within the Cr-Ni-V system that are typically used to assess the feasibility of the gradient compositions during path planning.

To achieve reliable predictions for path planning, thermodynamic databases that are validated over the entire composition range of interest are required. However, these are often restricting, as databases used for phase predictions are typically optimized for one principal element in a material system [16,17]. There are commercial, closed databases available that are intended for high entropy alloys; but these databases still do not include a complete description of ternary or higher order systems that exist within the entire material system [18].

Thermodynamic modeling of the Cr-Ni-V system has been performed by Choi et al. [13,14] and is currently the only available description of the ternary system with reasonable phase models. However, due to the lack of ternary experimental thermochemical data available in the literature for the Cr-Ni-V system, they used the binary CALPHAD modeling works of Cr-Ni by Lee et al. [19], Cr-V by Lee et al. [20], and Ni-V by Korb et al. [21] to model their ternary system and did not include ternary excess parameters for any phases except for the topologically close-packed (TCP) sigma phase [13,14]. Furthermore, Choi et al. [13,14] used the sublattice model (Cr, V)$_4$(Ni)$_8$(Cr, Ni, V)$_{18}$ to describe the sigma phase, which combines the Wyckoff positions of 2a, 8i$_1$, and 8j as listed in **Table 1**. However, recent CALPHAD modeling of the Ni-V binary system from Noori and Hallstedt [22] uses a three sublattice model (Cr, Ni, V)$_4$(Cr, Ni, V)$_{10}$(Cr, Ni, V)$_{16}$ that combines the 2a and 8i$_2$ Wyckoff positions with 12 coordination numbers and, sites 8i$_1$ and 8j with 14 coordination numbers. This sublattice model is more appropriate to use since the (Cr, Ni, V)$_4$(Cr, Ni, V)$_{10}$(Cr, Ni, V)$_{16}$ model has been shown by Joubert to better describe partitions of elements into different Wyckoff positions observed from experiments than the (Cr, V)$_4$(Ni)$_8$(Cr, Ni, V)$_{18}$ sublattice model [23], particularly for the Ni-V system [24], and it better represents the homogeneity range of sigma phase in the Cr-Ni-V system, thus, leading to more reliable phase predictions than those from the database in refs. [13,14].

The present work focused on remodeling the Cr-Ni-V system using the CALPHAD method through incorporating more recent experimental data and thermodynamic descriptions available in the literature. To incorporate ternary interaction parameters into the present database, DFT-based first-principles calculations were performed to calculate the formation energies of sigma phase using the $(Cr, Ni, V)_4(Cr, Ni, V)_{10}(Cr, Ni, V)_{16}$ sublattice model [25]. Additionally, DFT-based *ab initio* molecular dynamics (AIMD) simulations were used to predict the enthalpy of mixing ($\Delta H_{mix}$) of liquid, which was used to improve the modeling of liquid in the ternary system. This is important because Scheil prediction of phases in DED AM requires accurate modeling of liquid [26] as the behavior of the liquid affects the formation of solid phases during solidification. Scheil simulations predict phase formations under fast cooling, assuming no diffusion in the solid, and infinite mixing in the liquid, and have previously been shown to be an effective method for modeling phase formations in functionally graded AM materials [27].

To validate the model constructed for the present work and compare it with the most recent one for Cr-Ni-V (i.e., that by Choi et al. in refs. [13,14]), an FGM grading between Ni-20Cr and V was fabricated using DED AM. The phases within the FGM were experimentally characterized and compared with predictions from both thermodynamic databases and their feasibility maps [12,15], a concept developed by the present team, to assess the feasibility of the composition path grading through ternary composition space. The ability of these models to predict the experimentally observed formation of sigma phase in the Cr-Ni-V system was assessed. As Boron was found in the fabricated NiCr-V FGM and is frequently used in DED of Ti-6Al-4V for grain refinement [28], it was also incorporated into the present thermodynamic modeling to quantify its influence on the phases that formed in the DED process. The addition of B was an important consideration for understanding the effect of contaminants on AM fabrication.

## 2. Methodology

### 2.1. CALPHAD modeling of thermodynamics and associated input data

CALPHAD modeling of the thermodynamic properties of the ternary Cr-Ni-V system involves thermodynamically modeling the three pure elements, three binary systems, and the overall ternary system considered in the system. In the present work, thermodynamic models of pure elements were sourced from the Scientific Group Thermodata Europe (SGTE) database [29]. The Cr-Ni binary modeled by Tang et al. [30] was adopted as shown in **Figure S1**, because the remodeling by Hao et al. [31] used different models for pure elements and could not be combined with other binary systems considered in the present work. For the Ni-V binary system, Noori and Hallstedt [22] modeled $Ni_3V$ and $Ni_2V_7$ as non-stoichiometric phases which was used in the present work as shown in **Figure S2**. The model from Korb et al. [21] for Ni-V, which treats these compounds as stoichiometric phases, was used by Choi et al. [13,14] in their modeling of the Cr-Ni-V ternary system. The modeling of the Cr-V system by Ghosh et al. [32] was used and improved in the present work with the latest enthalpy of formation ($\Delta H_f$) predictions of the bcc phase by Gao et al. [33] through DFT-based first-principles calculations.

There are two classes of phases contained in the Cr-Ni-V system - solution phases with a single sublattice and compounds that contain multiple sublattices. These are typically differentiated by a phase's Wyckoff positions. The liquid phase is typically modeled with one-sublattice model unless there is strong short-range ordering. Using the Gibbs energy functions of pure elements (i.e., Cr, Ni, and V) taken from the SGTE database [29], the Gibbs energy functions of solution phases liquid, bcc, and fcc in the Cr-Ni-V system are described by the Redlich-Kister polynomial [34] as follows,

$$G_m^\alpha = x_{Cr}G_{Cr}^\alpha + x_{Ni}G_{Ni}^\alpha + x_V G_V^\alpha + RT(x_{Cr}lnx_{Cr} + x_{Ni}lnx_{Ni} + x_V lnx_V) \quad \text{Eq. 1}$$

$$+ x_{Cr}x_{Ni}\sum_{k=0} {}^kL_{Cr,Ni}(x_{Cr} - x_{Ni})^k + x_{Cr}x_V \sum_{k=0} {}^kL_{Cr,V}(x_{Cr} - x_V)^k$$

$$+ x_{Ni}x_V \sum_{k=0} {}^kL_{Ni,V}(x_{Ni} - x_V)^k + x_{Cr}x_{Ni}x_V(x_{Cr}{}^{Cr}L_{Cr,Ni,V}$$

$$+ x_{Ni}{}^{Ni}L_{Cr,Ni,V} + x_V{}^V L_{Cr,Ni,V})$$

where $x_{Cr}$, $x_{Ni}$, and $x_V$ are the mole fractions of Cr, Ni, and V in phase $\alpha$, respectively, $G_{Cr}^\alpha$, $G_{Ni}^\alpha$, and $G_V^\alpha$ represent the Gibbs energies of Cr, Ni, and V in phase $\alpha$ with respect to their standard element reference (SER) states at pressure $P = 1$ bar and temperature $T = 298.15$ K, R is the gas constant, ${}^kL_{i,j} = {}^ka_{i,j} + {}^kb_{i,j}T$ corresponds to the $k^{th}$ interaction parameter between $i$ and $j$ with ${}^ka_{i,j}$ and ${}^kb_{i,j}$ as model parameters, and ${}^iL_{Cr,Ni,V} = {}^ia_{Cr,Ni,V} + {}^ib_{Cr,Ni,V}T$ represents the ternary interaction for component $i$ with ${}^ia_{Cr,Ni,V}$ and ${}^ib_{Cr,Ni,V}$ as model parameters.

The Gibbs energy of compounds with multiple Wyckoff positions can be described by a sublattice model, with one sublattice for each type of Wyckoff site using the compound energy formalism [35],

$$G_{mf} = {}^0G_{mf} + RT \sum_t a^t \sum_j y_j^t lny_j^t + {}^EG_{mf} \quad \text{Eq. 2}$$

where ${}^0G_{mf} = \sum_{em_i}(\prod_t y_{em_i}^t {}^0G_{em_i})$ represents the contributions from Gibbs energy of each endmember with only one component in each sublattice, ${}^0G_{em_i} = \sum_t a^t {}^0G_{em_i}^t$, with $y_{em_i}^t$ and ${}^0G_{em_i}^t$ being the site fraction and Gibbs energy of the component in sublattice $t$ in the endmember $em_i$, respectively, and $a^t$ being the multiplicity of sublattice $t$. The second term corresponds to the ideal entropy of mixing summed over each sublattice. The last term, ${}^EG_{mf}$, represents the excess Gibbs energy that includes two types of contributions: the first being the interaction in one sublattice, where all other sublattices only contain one component (n), the second being the mixing simultaneously in two or more sublattices, where more than one sublattice contains two or more components. The present work focused on the first type of interaction among Cr, Ni, and V as follows,

$${}^E G_{mf} = \sum_t \prod_{s \neq t} y_i^s \sum_{i>j} \sum_j y_i^t y_j^t L_{i,j:l}^t \qquad \text{Eq. 3}$$

where $L_{i,j:l}^t$ denotes the interaction parameter between components $i$ and $j$ in the sublattice $t$ with all other sublattices containing only one component in each sublattice. As discussed above, the sigma phase is important in the present work. Since this phase has five Wyckoff positions, it is desirable to model it using five sublattices, which is being actively pursued by the team. However, in the present work, existing CALPHAD modeling of the Ni-V binary system [22] was adopted along with its three sublattice model, i.e., (Cr, Ni, V)$_4$(Cr, Ni, V)$_{10}$(Cr, Ni, V)$_{16}$.

Models for other binary compounds of CrNi$_2$, NiV$_3$, Ni$_3$V and Ni$_8$V that were considered in the present modeling were directly taken from Cr-Ni from Tang et al. [30] and Ni-V from Noori and Hallstedt [22]. No ternary interactions were added to these binary compounds due to their unimportant role in the present work.

### 2.1.1. Literature data for the Cr-V binary system

The Cr-V system was first remodeled using $\Delta H_f$, activity, and phase equilibria data. The accepted Cr-V binary system contains only two phases which are both solution phases: liquid and bcc. $\Delta H_f$ predictions for compositions in the Cr-V system at 0 K were also incorporated into the present models. Gao et al. [33] studied the ground-state structures at 0 K using a cluster expansion method fitted to DFT-based first-principles calculations. By considering 18 structures from the compounds in the Cr–TM (TM = transition metal) binary systems [36] they identified nine stable structures on the convex hull. Based on their work, these nine stable structures were used for the present CALPHAD modeling. The activities of Cr in the Cr-V system used in the present CALPHAD modeling were determined between 10.0 to 90.0 at. % V by Aldred and Myles [37] at 1550 K by vapor pressure measurements using the torsion-effusion method.

Four different experimental measurements were considered to determine phase equilibrium data for the Cr-V system. However, differences were observed in the liquidus-solidus equilibria measured in the literature, and thus, only data with similar reported values were used in the present modeling. Carlson et al. [38] found that the lowest solidus temperature of 2023 K occurred at 30.0 at.% V by noting the lowest temperature at which liquid was observed during optical pyrometry; Kubaschewski et al. [39] used empirical methods to estimate the lowest solidus temperature of 2068 K at around 75.0 at.% V; and the most recent data reported by Smith et al. [40] showed the minimum solidus temperature to be 2038 K at 20.0 at.% V, measured by optical pyrometry. The existence of a lowest solidus temperature is supported by the activity of Cr in Cr-V reported by Aldred et al. [37], which showed that the lowest solidus temperature was at 30.0 at.% V. As multiple measurements (i.e., Carlson et al. [38] and Smith et al. [40]), reported the lowest solidus temperature to occur between 20 – 30 at. % in the temperature range of 2023 – 2038 K, these were the data considered in the present work.

### 2.1.2. DFT-based first-principles calculations for liquid and sigma phases in Cr-Ni-V

In the present study, the DFT-based quasiharmonic approach is used to calculate $\Delta H_f$ for sigma, while AIMD calculations were used to obtain the enthalpy of mixing ($\Delta H_{mix}$) for liquid phase. This is due to the lack of experimental data in other ternary sigma and liquid phases in other modeling works. According to the DFT-based quasiharmonic approach, Helmholtz energy of a solid phase can be estimated as a function of temperature, $T$, and volume, $V$, which corresponds to the Gibbs energy under zero external pressure [41],

$$F(V,T) = E_0(V) + F_{vib}(V,T) + F_{el}(V,T) \qquad \text{Eq. 4}$$

where the Helmholtz energy, $F$, includes the static energy at 0 K without the zero-point vibrational contribution, $E_0(V)$, the contribution from lattice vibrations, $F_{vib}$, and the contribution from thermal electrons, $F_{el}$. The equilibrium volume at each temperature was determined by finding the minimum of $F$, i.e., when the pressure, $P = -\frac{\partial F}{\partial V} = 0\ Pa$.

DFT-based calculations were used to predict the energy versus volume (E-V) curve for each phase (or endmember) at 0 K. The resulting data points were then fitted using a four-parameter Birch-Murnaghan (BM4) equation of state (EOS) [41],

$$E_0(V) = k_1 + k_2 V^{-2/3} + k_3 V^{-4/3} + k_4 V^{-2} \qquad \text{Eq. 5}$$

with fitting parameters $k_1$, $k_2$, $k_3$, and $k_4$. The equilibrium properties at $P = 0$ GPa from this EOS include the equilibrium energy, $E_0$, volume, $V_0$, bulk modulus, $B_0$, and derivative of the bulk modulus with respect to pressure, $B'$. The contribution of lattice vibrations, $F_{vib}$, was determined using the phonon density of states (pDOS) [42],

$$F_{vib}(T,V) = k_B T \int_0^\infty \ln\left[2\sinh\frac{\hbar\omega}{2k_B T}\right] g(\omega)\ d\omega \qquad \text{Eq. 6}$$

where $g(\omega)$ is the pDOS as a function of $V$ and frequency, $\omega$ [42], and $F_{el}$ was obtained using Mermin statistics [42].

For the liquid phase, AIMD was used to determine its $\Delta H_{mix}$ with composition $Cr_xNi_yV_{1-x-y}$ at 2700 K, calculated by the function,

$$\Delta H_{mix} = H_{mixture} - \sum c_i H_i \qquad \text{Eq. 7}$$

where $H_{mixture}$ is the enthalpy of Cr$_x$Ni$_y$V$_{1-x-y}$ calculated by AIMD, $c_i$ the mole fraction of the component $i$, and $H_i$ the enthalpy of pure elements Cr, Ni, and V at 2700 K.

In the present work, all DFT-based first-principles and AIMD calculations were conducted using the Vienna *ab initio* Simulation Package (VASP) [43]. The projector augmented wave (PAW) method [44] was used to describe the electron-ion interaction, while the generalized gradient approximation (GGA) by Perdew, Burke, and Ernzerhof (PBE) [45] was used to describe the exchange-correlation functional. For AIMD simulations, an NVT ensemble was used with the total number of atoms in the supercell, N, set to 108. A *k*-point mesh was set as a single Γ point 1×1×1, and the 280 eV was set as cutoff energy. A fixed temperature of 2700 K was used for AIMD simulations to ensure that all the six compositions were in the liquid state, and the Nose-Hoover thermostat was used to regulate the temperature [46,47].

To estimate first-principles thermodynamics of the Cr-Ni-V system, the reference states of pure elements bcc Cr, fcc Ni, and bcc V were used, where ferromagnetic spin-polarization are considered for the Cr and Ni atoms. The sigma phase was modeled using three sublattices consisting of 27 endmembers as opposed to the sublattice model from refs. [13,14]. **Table 2** summarizes the settings for DFT-based first-principles and AIMD calculations for each compound or pure element, including the information of total atoms in the supercells, k-point meshes used for structural relaxations and final static calculations (denoted by DFT), and k-point meshes for AIMD calculations. Spin-polarization was used in all the calculations. For modeling of the liquid phase, AIMD simulations were carried out for six compositions: Cr$_{108}$, Cr$_{27}$Ni$_{27}$V$_{54}$, Cr$_{27}$Ni$_{54}$V$_{27}$, Cr$_{54}$Ni$_{27}$V$_{27}$, Ni$_{108}$, and V$_{108}$.

### 2.1.3. Phase equilibrium data in Cr-Ni-V

For modeling of the ternary parameters, experimental phase equilibria measurements between bcc, fcc, and sigma phases were taken from Kodentzov et al. [48] which were collected using scanning electron microscope (SEM) imaging and microprobe analysis on samples annealed at 1275 K and 1425 K. At 1275 K, composition boundaries, in at%, between the sigma phase and bcc were identified to extend from 0.0Cr-25.8Ni-74.2V to 5.3Cr-11.0Ni-83.7V for the sigma phase boundary and 0.0Cr-12.8Ni-87.2V to 3.8Cr-27.4Ni-68.8V for the bcc boundary. For the two-phase sigma and fcc region, the sigma phase boundary ranged from 3.3Cr-44.4Ni-52.3V to 31.5Cr-39.6Ni-28.9V, and fcc extended from 3.0Cr-59.0Ni-38.0V to 20.6Cr-56.2Ni-23.2V.

At 1425 K, the sigma phase boundary extended between 7.3Cr-28.9Ni-63.8V to 42.3Cr-35.7Ni-22.0V and the bcc from 2.3Cr-16.4Ni-71.3V to 56.1Cr-24.2Ni-19.7V. The fcc boundary between fcc and sigma phase ranged from 11.1Cr-5.4Ni-33.5V to 16.3Cr-57.3Ni-26.4V, while the sigma boundary between the two phases ranged from 18.8Cr-41.0Ni-40.2V to 26.6Cr-39.4Ni-34.0V. The boundary between bcc and fcc was determined through measuring a tie line between the bcc and fcc phase, which was determined to have a composition of 58.4Cr-26.1Ni-15.5V on the bcc boundary and at 32.7Cr-49.4Ni-15.9V on the fcc boundary.

Additional experimental data by Singh et al. [49] and Malhotra et al. [50] were used to model ternary parameters in the present database. Using scanning electron microscope (SEM) imaging, energy dispersive spectroscopy (EDS), and X-ray diffraction (XRD), Singh et al. [50] measured phase equilibria between bcc, fcc, and sigma at 1373 K. Malhotra et al. [50] measured phase equilibria based on XRD. Singh et al. [49] and Malhotra et al. [50] showed varying results on the tie line edges between the fcc and sigma phases. On the fcc boundary, the composition from Singh et al. was reported to be 9.8Cr-62.3Ni-27.9V [49] while Malhotra et al. reported 15.3Cr-61.9Ni-22.8V [50]. At the sigma boundary, the tie line edge was found to be at a composition of 4.9Cr-28.3Ni-66.8V from Singh et al. [49] while Malhotra et al. reported the composition as 10.4Cr-29.1Ni-60.5V [50]. Additionally, only Singh et al. [49] provided data on the sigma boundary between sigma and fcc, which extend from 24.5Cr-40.6Ni-34.9V to 29.2Cr-40.0Ni-30.8V. The bcc boundary between the sigma and bcc phases was reported to range from 6.3Cr-10.0Ni-83.7V to 46.6Cr-11.8Ni-41.6V in the work by Singh et al. [49]. All these experimental data are compared to the ternary phase diagram predictions between the two databases at the given temperatures in **Figure 6**.

## 2.2. Feasibility mapping

Feasibility maps, a tool developed by the team for use in AM [51], were constructed to assess the feasible build space within the ternary Cr-Ni-V alloy system as shown in **Figure 2**. These maps assess the feasibility of compositions and gradient pathways fabricated via AM based on the amount(s) of deleterious phase(s) that form as predicted by both Scheil and equilibrium calculations. These equilibrium calculations, shown in **Figure 1**(a), provide estimates on phase formations under infinitely slow cooling, which could be approached during in-process thermal cycling with each deposition of a new layer in the build, or post-process heat treatment. Conversely, Scheil calculations, shown in **Figure 1**(b), provide estimates on phase formation from the melt under infinitely rapid solidification, which has been shown to be suitable for predicting phases present in as-built AM components [27].

In the Cr-Ni-V, the feasibility of a composition is determined based on if the total amount of sigma, is below (feasible) or above (infeasible) a threshold that is assumed to lead to cracking [12]. The feasibility maps used equilibrium calculations ranging between 1000 K, which is approximately 2/3 of the lowest predicted solidus temperature along the linear composition gradient path between NiCr and V, to just above the liquidus temperature, 2000 K, in 10 K increments. This temperature range was used to estimate bounds for capturing kinetically driven phase formations in AM processes that could potentially induce solid state phase transformations within this temperature range [52]. Scheil simulations were performed with a starting temperature above the liquidus, to model rapid solidification from a liquid melt. Together, these two predictions provide the bounds of two extreme cases between which phase transformations in the far from equilibrium DED AM process are expected to fall. These calculations were used to determine whether amounts of deleterious phase exceeded the tolerated amount across all compositions in the Cr-Ni-V ternary.

**2.3. Experimental methods**

DED AM was used to fabricate an FGM grading between Ni-20Cr and 59 wt. % V (balance Ni-20Cr) to validate the thermodynamic database developed in the present work. Powders of Ni-20Cr (American elements, California), (herein referred to as NiCr) and V (American Elements, California), both sieved to a -100/+325 mesh size were used. The FGM was fabricated using a DED AM system (RPM, Inc., Model 222, South Dakota) in an Ar environment (<10 ppm $O_2$) equipped with a YAG laser that was operated using the parameters given in **Table 3**. Note that the processing parameters varied throughout the builds to accommodate the varying melting temperature of the compositions being deposited as a function of height. The resulting build was approximately 25.4 mm in diameter and 76.2 mm in height.

An as-built sample was prepared for evaluation by sectioning a pillar, mounting in epoxy resin, then polishing using standard metallographic techniques with a final polish of 0.05 μm alumina slurry. EDS was performed in an SEM (Thermo-Scientific, Apreo S Low Vac,

Massachusetts) equipped with a silicon drift detector (Oxford Instruments, Ultim Max silicon drift detector, Massachusetts) to measure local compositions within the sample. Local phase identification was performed using electron backscatter diffraction (EBSD, Oxford Instruments, Symmetry detector, Massachusetts). Electron Probe Microanalysis (EPMA, Cameca SX-5, Wisconsin) equipped with a $LaB_6$ electron source was used in selected areas to determine the compositions of the boride particles found within the FGM build.

## 3. Results and discussion

### 3.1. Computational results

#### 3.1.1. Thermodynamic properties by first-principles calculations

**Table 4** summarizes the DFT-predicted values of $\Delta H_f$ for the endmembers in sigma phase based on the sublattice model of $(Cr, Ni, V)_4(Cr, Ni, V)_{10}(Cr, Ni, V)_{16}$ used in the present CALPHAD modeling. The DFT calculations show that the $\Delta H_f$ values for the endmembers containing single elements such as $(Cr)_4(Cr)_{10}(Cr)_{16}$, $(Ni)_4(Ni)_{10}(Ni)_{16}$, and $(V)_4(V)_{10}(V)_{16}$ are positive and therefore not stable. Similarly, the $\Delta H_f$ values for all endmembers in the Cr-V and Cr-Ni binary systems are also positive and not stable, including $(Cr)_4(Cr)_{10}(V)_{16}$, $(Cr)_4(V)_{10}(Cr)_{16}$, $(Cr)_4(V)_{10}(V)_{16}$, $(V)_4(V)_{10}(Cr)_{16}$, $(V)_4(Cr)_{10}(Cr)_{16}$, $(V)_4(Cr)_{10}(V)_{16}$, $(Cr)_4(Cr)_{10}(Ni)_{16}$, $(Cr)_4(Ni)_{10}(Cr)_{16}$, $(Cr)_4(Ni)_{10}(Ni)_{16}$, $(Ni)_4(Ni)_{10}(Cr)_{16}$, $(Ni)_4(Cr)_{10}(Cr)_{16}$, and $(Ni)_4(Cr)_{10}(Ni)_{16}$, as listed in **Table 4**. However, some endmembers in Ni-V, such as $(V)_4(V)_{10}(Ni)_{16}$, $(V)_4(Ni)_{10}(Ni)_{16}$, $(Ni)_4(V)_{10}(V)_{16}$, $(Ni)_4(V)_{10}(Ni)_{16}$, $(Ni)_4(Ni)_{10}(V)_{16}$ have negative $\Delta H_f$ values. In addition, some endmembers in Cr-Ni-V, such as $(Ni)_4(Cr)_{10}(V)_{16}$, $(Ni)_4(V)_{10}(Cr)_{16}$, $(V)_4(Cr)_{10}(Ni)_{16}$ have negative $\Delta H_f$ values.

For comparison to the present modeling, predictions for activity values at 1550 K of the Cr-V binary from the present work were plotted in **Figure 3**(a) with respect to those calculated by the modeling from Ghosh et al. [32] and experimental data from Aldred et al. [37]. As shown, the activity from both the present modeling and that from Ghosh et al. [32] match well with experimental data in the literature. However, the present modeling shows better predictions for experimental data between 10.0 to 40.0 at. % V, while Ghosh's modeling [32] provides a better match between 60.0 to 90.0 at. % V. On average, the difference between the present modeling results for activity and the experimental data is 0.0077, while results from Ghosh's modeling differs by 0.0087 [32].

$\Delta H_f$ were also compared between DFT calculations [33], the present modeling, and that developed by Ghosh et al. [32] at 298 K in **Figure 3**(b). The calculated $\Delta H_f$ reported by Gao et al. [33] decreases as the amount of V content increases between 0.0 to 44.0 at. %, then reaches a minimum value of -7.42 kJ/mol-atom. The present modeling shows good agreement with the DFT calculations from Gao et al. [33], with an average difference of approximately 0.41 kJ/mol-atom, while the difference between Ghosh's modeling [32] and the DFT calculations is about 3.91 kJ/mol-atom. Thus, the present modeling demonstrates good agreement with thermochemical data for the Cr-V system, which includes the activity values of Cr at 1550 K, and $\Delta H_f$ data.

**Figure 4** shows the $\Delta H_{mix}$ values of liquid at 2700 K from both the present modeling and that of Choi et al. [13,14], compared to the present AIMD results at 2700 K. It is shown that the present modeling predicts the lowest $\Delta H_{mix}$ (-25,000 J/mol-atom) at around 33.3Cr-33.3Ni-33.3V. In contrast, the modeling by Choi et al. [13,14] gives the lowest $\Delta H_{mix}$ (-13,000 J/mol-atom), at around 0Cr-50.0Ni-50.0V. The present AIMD simulations predict -24,039.5 J/mol-atom at 25.0Cr-50.0Ni-25.0V, -22,706.7 J/mol-atom at 50.0Cr-25.0Ni-25.0V, and -22,758.0 J/mol-atom at 25.0Cr-25.0Ni-50.0V. Therefore, the present modeling shows a better agreement with the AIMD results than the modeling from Choi et al. [13,14], with an average difference around 1,275 J/mol-atom versus a difference of 16,763 J/mol-atom between Choi's work [13,14] and the AIMD results. These differences in $\Delta H_{mix}$ are attributed to the lack of ternary interactions for liquid in the modeling by Choi et al. [13,14]. Because the present CALPHAD modeling more accurately describes the liquid phase than that by Choi et al. [13,14], better predictions of phase stability under different cooling rates are achieved, as described further in Section 4.2.

### 3.1.2. Thermodynamic modeling of Cr-V and Cr-Ni-V

**Figure 5** presents the Cr-V phase diagram predicted by both the present CALPHAD modeling and that by Ghosh et al. [32], compared to experimental data by Carlson et al. [38] and Smith et al. [40]. The phase diagram from the present modeling exhibits a local minimum in the liquidus at 32.5 at. % V at a temperature of 2026 K. The phase diagram from the present work agrees well with the data from Smith et al. [40] at 20.0 at. % V while between 30.0 to 80.0 at. % V, the model shows a better match with data from Carlson et al. [38]. The average difference between the present work's phase boundaries and data from Smith et al. [40] and from Carlson et al. [38] is around 12 K and 11 K, respectively. In contrast, the phase diagram from Ghosh's modeling [32] agrees well with data from Smith et al. [40], but not with those from Carlson et al. [38]. The average difference between the phase diagram from Ghosh's modeling [32] and experimental data from Smith et al. [40] and Carlson et al. [38] is around 6 K and 23 K, respectively. Thus, the phase diagram from the present work matches well with both sets of experimental data, while Ghosh's modeling [32] provides a better match only with Smith et al. [40].

**Figure 6** shows the predicted isothermal sections of Cr-Ni-V from the present modeling and Choi et al.'s modeling [13,14] at (a, b) 1275 K, (c, d) 1373 K, and (e, f) 1425 K compared to experimental data from Kodentzov et al. [48] at 1275 K and 1425 K, and Singh et al. [49] and Malhotra et al. [50] at 1373 K. As shown in **Figure 6**(a) and (b), both modeling works provide a good match with experimental data at 1275 K for the boundary between the sigma phase and bcc, which ranges from 0.0Cr-25.8Ni-74.2V to 5.3Cr-11.0Ni-83.7V on the sigma side and 0.0Cr-12.8Ni-87.2V to 3.8Cr-27.4Ni-68.8V on the bcc side. Regarding the two-phase sigma and fcc region, the present work provides a better match to experimental data on the sigma boundary that ranges from 3.3Cr-44.4Ni-52.3V to 31.5Cr-38.9Ni-29.6V and the fcc phase, ranging from 3.3Cr-44.4Ni-52.3V to 31.5Cr-39.6Ni-28.9V. In contrast, the modeling by Choi et al. [13,14] depicts the sigma phase boundary to be between 3.3Cr-43.0Ni-53.7V to 31.5Cr-34.5Ni-34.0V and does not agree with experimental data which ranges from 3.3Cr-44.4Ni-52.3V to 31.5Cr-39.6Ni-26.9V.

**Figure 6** (c-d) give the isothermal section at 1373 K for Cr-Ni-V from the present modeling and Choi et al.'s modeling [13,14], overlayed with experimental data from Singh et al. [49] and Malhotra et al. [50]. Although experimental data from Singh et al. [49] and Malhotra et al. [50] exhibit scatter along the fcc boundary that extends from 9.8Cr-62.3Ni-27.9V to 15.3Cr-61.9Ni-22.8V between the sigma and fcc phases, the sigma phase boundary between sigma and bcc which spans from 4.9Cr-28.3Ni-66.8V to 10.4Cr-29.1Ni-60.5V from both models match well with experimental data. However, Choi et al.'s modeling [13,14] predicts a sigma boundary between the two-phase sigma and bcc region to occur at 25.5Cr-29.7Ni-44.8V and does not align with the experimental data from Singh et al. [49] observed at 25.5Cr-32.2Ni-42.3V. In contrast, the present modeling provides a boundary prediction at a composition of 25.5Cr-31.5Ni-43.0V which better matches experimental data on the sigma boundary. Additionally, the present modeling provides a boundary for sigma that ranges from 24.5Cr-39.7Ni-35.8V to 29.2Cr-39.2Ni-31.6V between the sigma and bcc phases, which better match the experimental data from compositions at 24.5Cr-40.6Ni-34.9V and 29.2Cr-40.0Ni-30.8V, compared to Choi et al.'s modeling [13,14] which predicts a boundary between 24.5Cr-36.5Ni-39.0V and 29.2Cr-35.3Ni-35.5V.

**Figure 6**(e-f) provide a comparison between isothermal sections at 1425 K for the two databases overlayed with experimental data from Kodentzov et al. [48] which shows good agreement between the present modeling and experiments. The present modeling has differences of less than 1.5 at. % Cr and 1.5 at. % Ni between experiments and predictions along all boundaries, except for at the sigma boundary at a composition predicted to be 39.5Cr-36.5Ni-24.0V, while experimentally it was observed at 42.5Cr-36.5Ni-21.0V [48]. In contrast, Choi et al.'s modeling [13,14] matches well with the experimental data for the boundary of the sigma phase between the two-phase sigma and bcc region, where the sigma boundary extends from 7.3Cr-28.9Ni-63.8V to 34.9Cr-30.7Ni-34.4V and the bcc boundary extends from 2.3Cr-16.4Ni-81.3V to 51.1Cr-18.9Ni-30.0V. However, Choi et al.'s modeling [13,14] does not align with the experimentally determined sigma boundary at 42.3Cr-35.7Ni-22.0V or the bcc boundary experimentally observed at 56.1Cr-24.2Ni-19.7V [48].

Additionally, Choi et al.'s modeling [13,14] matches well with experimental data for the fcc and sigma boundaries of the two-phase region contained between them. The fcc boundary is predicted to occur between 11.1Cr-55.4Ni-33.5V to 16.3Cr-57.3Ni-26.4V while the sigma boundary ranges from 18.8Cr-41.0Ni-40.2V to 26.6Cr-39.4Ni-34.0V. Furthermore, Choi et al.'s modeling [13,14] of the bcc boundary 58.4Cr-27.8Ni-13.8V and fcc boundary 32.7Cr-48.4Ni-18.9V along a predicted tie line did not match those compositions in the experimentally observed bcc composition, 58.4Cr-26.1Ni-15.5V, and fcc composition, 32.7Cr-49.4Ni-17.9V. These results show that the present modeling is in better agreement with experimental data than the modeling from Choi et al. [13,14], particularly for the sigma boundary between the two-phase sigma and fcc region at 1275 K and 1425 K.

### 3.2. Experimental results

To demonstrate the importance of a validated and accurate database for predicting phase formation in alloys, additional experimental data were collected in the present work to validate the Cr-Ni-V model using the fabricated NiCr-V FGM. To determine the phases along the linear NiCr to V composition pathway, EDS area scans and EBSD data were collected across the height of the FGM. EDS and EBSD analysis focused on 11 distinct compositions along the FGM that ranged from 3 to 59 wt.% V, as shown in **Figure 7**. From here on, compositions will be defined in wt.% V with the understanding that the balance is NiCr.

Faceted particles, which will be discussed later, were present in regions with 16 – 59 wt% V and were likely a result of powder contamination with boron. Small round particles were also present throughout the areas and were found to be carbides. To focus on the phases present in just the Cr-Ni-V ternary, the compositions of the areas surrounding these particles, referred to as the matrix, were analyzed and used for comparison to computational phase predictions. EDS area scans of approximately 65 μm x 45 μm in size were sub-sectioned into five 13 μm x 45 μm areas, from which composition data were averaged and used to calculate the standard deviation of the overall scan area. The compositions from this analysis are shown in **Figure 8** in comparison with those for an ideal linear composition gradient between NiCr and V and were found to be in alignment with each other.

The EBSD scans were analyzed for fcc, bcc, and intermetallic compounds of $CrNi_2$, sigma, $NiV_3$, and $Ni_2V_7$ as they are compounds within the Cr-Ni-V ternary system. Average area fractions and standard deviations of phases for each layer composition were calculated using the same method above on the scans for composition analysis. The amount of zero-solution for each area was added to the positive direction of the error bars to account for the uncertainty in what phases were present in those regions, and the resulting phase fractions as a function of composition in the FGM are compared with computational predictions in **Figure 1**. Considering only the matrix phase(s), EBSD analysis indicated that only fcc was present from compositions of 3 wt.% V up to 25 wt.% V, a two-phase mixture of sigma and fcc was present between 34 and 42 wt% V, and for 52 and 54 wt% V, only sigma phase was observed. At a composition of 59 wt.% V, primarily sigma phase was present with a small amount of bcc phase, as shown in the phase maps in **Figure 7**.

Additional composition and phase analyses were performed on the small round particles that appeared between 11 and 34 wt.% V and the faceted particles that appeared between 16 and 59 wt.% V. EDS area scans indicated that the round particles were enriched in V and C, as shown in **Figure 7**, and EBSD analysis identified these particles to be fcc VC. EDS area scans of the irregular particles and interdendrite areas of the 3 – 16 wt% V locations indicated that these features were enriched in B as shown in **Figure 7**. EPMA analysis confirmed the composition of the particles to be approximately that of MB and $M_3B_2$ compounds. The presence of borides was further confirmed by EBSD phase analysis, which identified these particles to have crystal structure that correspond to MB and $M_3B_2$ phases, where M indicates a mixture of Cr and V [53]. As B should not have been present in this FGM, it is likely that the powders were contaminated during pre-processing (e.g., sieving or storage) or in-process (e.g., due to residual boron in the feedlines from a previous build).

While not the focus of this study, to confirm the formation of borides in this Cr-Ni-V system contaminated with boron, the borides that experimentally appeared in the sample were modeled using the CALPHAD method and were incorporated into the modeling for the present work. The model adopted the B-containing binary phase diagrams from the latest publications on Cr-B from Tojo et al. [54], Ni-B from Oikawa et al. [55], and V-B from da Silva et al. [56]. The $\Delta H_f$ of MB and $M_3B_2$ were calculated from both DFT [57,58] and the machine learning (ML) model from SIPFENN [59] which were trained based on OQMD data [57,58] with a mean absolute error (MAE) of 41.9 meV/atom. Based on the modeling, the lowest $\Delta H_f$ appears at $(V)_{0.5}(B)_{0.5}$ for MB and $(Cr_{0.25}V_{0.75})_{0.6}(B)_{0.4}$ for $M_3B_2$, indicating there are two stable ternary borides in the B-Cr-Ni-V system, MB and $M_3B_2$.

With the updated model for the B-Cr-Ni-V system, Scheil simulations were performed to thermodynamically explain the formation of borides found in the NiCr-V FGM and to understand the effect of B on phase formation in the sample. The overall B content was estimated by using weighted fractions between the particle compositions and the matrix. Particle compositions were calculated using stoichiometry of the MB and $M_3B_2$ compounds and EDS data for the fractions of Cr and V contained within the particles then were weighted according to the phase fractions of the phases observed from EBSD analysis. As shown in **Table 5**, the computationally predicted phase fractions from Scheil simulations performed on these B-containing compositions aligned with the experimentally observed phase results for the overall area compositions, demonstrating the accuracy of the database modeling.

To understand the influence of B on the phase formation in the NiCr-V system, comparisons between experimentally observed phase fractions were made between the matrix area and the entire overall area (including the borides) by calculating the relative ratio of fcc, sigma, and bcc phases in these areas, as given in **Table 6**. This calculation was performed to remove the effect of borides on the phase fraction amounts observed in the overall area. These relative phase fractions were then compared to the experimentally observed phase fractions in the matrix and were found to fall within the error calculations of the experimentally observed matrix phase(s) in the NiCr-V FGM. Thus, it was concluded that the presence of borides in the build did not have a distinguishable effect on the phase formation, which is reasonable considering that boride particles have significantly higher melting temperatures (~2300 K for $V_3B_2$ and 2500 K for VB) [60], resulting in those compounds solidifying first, and leaving behind a melt pool composition made up of Cr-Ni-V from which solidification would proceed. It is hypothesized that this difference in melting temperatures would lead to the borides solidifying out of the melt prior to solidification of any other phases, such that they would not participate in the rest of the subsequent solidification of phases from the remaining melt.

### 3.3. Comparison of experimental results with computational predictions

Thermodynamic calculations were performed using the present model and the model developed by Choi et al. [13,14] to predict phases as a function of composition in the Cr-Ni-V system, with experimental validation provided along the linear gradient path between NiCr and V. A comparison of the phases predicted by the two databases is shown in **Figure 1**, along with the experimentally collected phase data for the 11 compositions that were measured along the height of the FGM. **Figure 1**(a) provides the phases predicted by equilibrium calculations. As shown, the equilibrium calculations suggest a transition from fcc to sigma as V content reached 20 wt. % for the present model and 18 wt. % in that by Choi et al. [13,14], then a transition from sigma to bcc when the V content in the sample increases above 65 wt% from the present model and 63 wt. % for Choi et al.'s model [13,14]. Additionally, the equilibrium calculations using the present database predicts the $CrNi_2$ intermetallic phase to be present near the NiCr-rich end and $NiV_3$ near V-rich end, and Choi et al.'s database [13,14] predicts $Ni_2V_7$ to be present in the V-rich end along the gradient path, none of which were experimentally observed.

The differences in the equilibrium predictions between the two thermodynamic models is due to the non-stoichiometric phase descriptions for $CrNi_2$ and $Ni_3V$ that were incorporated into the present model, allowing for the $CrNi_2$ and $Ni_3V$ phases to form across a wider composition range than that in Choi et al.'s model [13,14]. The updated modeling in the present work features a modified sublattice model compared to Choi et al.'s [13,14], including the use of a $(Cr, Ni, V)_1(Cr, Ni)_2$ model instead of $(V)_1(Ni)_2$ to incorporate Cr into the $CrNi_2$ phase, which is the $VNi_2$ phase in Choi et al. [13,14] which is not predicted to form here. Additionally, the $(Ni, V)_1(Ni, V)_3$ sublattice model is used instead of the $(Ni)_2(V)_7$ sublattice in Choi et al. [21], resulting in a wider composition range for the $NiV_3$ phase to appear in the Ni-V system. With these differences, the present modeling predicts the formation of fcc, sigma, and bcc, and intermetallic phases $CrNi_2$ and $NiV_3$, while Choi et al.'s modeling [13,14] predicts fcc, sigma, and $Ni_2V_7$ phases to form across the same composition range.

While the equilibrium phase predictions at 1000 K were used to represent the lowest temperature threshold at which solid state phase transformations were assumed to be kinetically feasible during the FGM fabrication via DED AM, the intermetallic $CrNi_2$ phase that is predicted to form in the present database along the sample did not experimentally appear. Although previous work has suggested that kinetically driven phase transformations are possible from the solidus temperature down to approximately 1000 K for another Ni-Cr alloy [52], the experimental results here suggest that it is possible that the kinetically induced solid-state phase transformations may occur at temperatures higher than 1000 K in AM processing or may not have sufficient time between thermal cycles to occur. The Scheil solidification simulations were found to be a better predictor for phases formed in the DED AM deposition of the NiCr-V FGM, agreeing with our previous findings [27].

As shown in **Figure 1**(b), Scheil solidification simulations for both thermodynamic databases predict a smooth transition from fcc at the NiCr-rich end of the FGM, to sigma phase with the introduction of V, then to bcc at the V-rich end of the gradient, with no additional intermetallic phases predicted, which is in general agreement with the experimentally observed phases along the NiCr-V FGM. The present model aligns better with the experimental results than those from Choi et al. [13,14] as almost all the predictions from the present database fall within the error bounds of the experimentally observed results, while all predictions except for those in the NiCr-rich end from Choi et al. [13,14] are shown to deviate from experimental observations, with even more pronounced differences at higher V contents across the gradient.

Although both CALPHAD models predicted the same phases to appear in the matrix compositions investigated, they differ in the phase fractions of each phase predicted. For example, the present model consistently predicts more sigma phase to form than in Choi et al.'s model [13,14] between 32 to 71 wt% V. The difference in Scheil simulation predictions between the present model and that from Choi et al.'s [13,14] is a result of the difference in modeling of liquid in the Cr-Ni-V ternary system. The ternary interaction parameters dictate the liquidus and solidus temperatures and affect the Scheil simulations significantly. Thus, the corrections made to the liquid phase in the present model improve the accuracy of the Scheil predictions.

### 3.4. Feasibility mapping of FGMs and crack location

Experimental data were used for validation of the feasibility maps [12,15] constructed of the two CALPHAD models for evaluation of feasible and infeasible compositions within the Cr-Ni-V ternary composition space. Special attention was given to the regions designated as Scheil infeasible due to the alignment of Scheil predictions with experimentally observed phases present along the FGM. As shown in **Figure 2**, the two databases predict varying infeasible and feasible areas within the build space, with the present database predicting a wider range of infeasible compositions near the Cr-rich corner from both Scheil and equilibrium predictions.

For the linear gradient between NiCr and V, both feasibility maps correctly predict the infeasible build compositions between 34 – 59 wt% V using the Scheil simulations, due to having higher than the allowed 0.10 mole fraction detrimental (sigma) phase. These compositions are depicted in **Figure 2**, and are contained within the regions marked in dark red as equilibrium and Scheil infeasible. With experimental validation of this area confirming it to be definitively infeasible due to having higher than the allowable threshold of sigma phase, a new potential pathway was plotted along the feasibility maps to grade around regions predicted to form these high amounts of deleterious phase by avoiding the regions predicted as to be (equilibrium and) Scheil infeasible.

However, due to different feasibility predictions between the feasibility maps of the two models, care must be taken in evaluating a CALPHAD model's ability to predict accurate amounts of phases in a material system. **Figure 2** presents a single alternative gradient pathway drawn on a feasibility map of the present model and the model by Choi et al. [13,14]. The path is predicted to be feasible using the modeling by Choi et al. [13,14] in **Figure 2**(a), and is not feasible when drawn across the feasibility map from the present model, shown in **Figure 2**(b). Because the present model is more suitable for AM phase predictions due to the improvements in the liquid phases and updated binary models, these results indicate that it would be necessary to grade further out into Cr-rich compositions to avoid high amounts of deleterious sigma phase from forming than suggested by the prediction using Choi et al.'s thermodynamic database [13,14]. These differences in feasible build space between the two CALPHAD model predictions for the Cr-Ni-V system demonstrate the importance of using experimentally validated, accurate liquid phase modeling in thermodynamic modeling for informed path planning in additively manufactured FGMs.

Cracking was also experimentally observed in the NiCr-V FGM gradient around 52 wt.% V, as noted on the feasibility maps in **Figure 2**. At this composition, both experiments and simulations from the present database show 100% brittle sigma phase, which was accurately predicted in both feasibility maps to be infeasible. While it is unknown what the exact cause of cracking is in this FGM due to its fabrication resulting in a nonlinear gradient as a function of height, it is thought that the sigma phase is too brittle to withstand the in-situ thermal cycles in addition to the stresses that arise from the sudden composition jumps along the build height. Regardless, it is undesirable to grade into regions containing this much sigma phase, and thus, a linear composition gradient between NiCr and V is an undesirable path through which to grade an FGM, in joining, for example, stainless steel and Ti-6Al-4V.

## 4. Conclusions

The present work used thermodynamic data from DFT-based first-principles and AIMD simulations, and recent experimental data from literature to remodel the Cr-V and Cr-Ni-V systems. The Cr-V system was remodeled using the DFT calculations from Gao et al. [33], while the ternary liquid was remodeled based on the AIMD simulations. To validate the present database for the ternary Cr-Ni-V system, an FGM grading from 0 wt. % V to 59 wt. % V (balance NiCr) was fabricated using DED AM. The compositions and phases of the FGM were characterized using EDS and EBSD. The experimentally identified phases as a function of composition were compared to equilibrium and Scheil solidification simulations using an existing thermodynamic database in the literature as well as the present database. The key findings are as follows:

- Improvements were made to the Cr-Ni-V thermodynamic modeling through the addition of accurate sublattice models for binary systems and the addition of ternary liquid parameters with the inputs from the present AIMD simulations, making the present thermodynamic model more suitable for predicting phases in an additively manufactured FGM over other Cr-Ni-V models.

- Scheil simulations performed using the present CALPHAD model accurately predicted the experimentally observed phases along the NiCr-V FGM. The present CALPHAD modeling and that by Choi et al. [13,14] predicted the same phases to appear along the sample height as a function of composition, while the present database more accurately captured the phase fractions of fcc, sigma, and bcc present at a given composition.

- A linear gradient from Ni-20Cr to V resulted in the formation of regions containing 100% sigma phase and is therefore an undesirable gradient pathway through the Cr-Ni-V ternary composition space.

- Boron contamination led to the formation of MB and $M_3B_2$ borides and while they do not have a significant influence on the phases that formed in the NiCr-V FGM, the observation of borides highlights the importance of identifying and controlling the amount of trace elements present in the AM process.

**Declaration of competing interest**

The authors declare that they have no known competing financial interests or personal relationships that could have appeared to influence the work reported in this paper.

**Data availability**

All relevant data are available from the authors.

**Acknowledgements**

B. Tonyali is supported by the National Defense Science and Engineering Graduate (NDSEG) Fellowship. This material is based upon work supported by the National Science Foundation under grant No. CMMI-2050069. Part of this research was carried out at the Jet Propulsion Laboratory, California Institute of Technology, under a contract with the National Aeronautics and Space Administration (80NM0018D0004). First-principles calculations were performed partially on the Roar supercomputer at the Pennsylvania State University's Institute for Computational and Data Sciences (ICDS), partially on the resources of the National Energy Research Scientific Computing Center (NERSC) supported by the U.S. DOE Office of Science User Facility operated under Contract No. DE-AC02-05CH11231, and partially on the resources of the Extreme Science and Engineering Discovery Environment (XSEDE) supported by NSF with Grant No. ACI-1548562.

# Figures

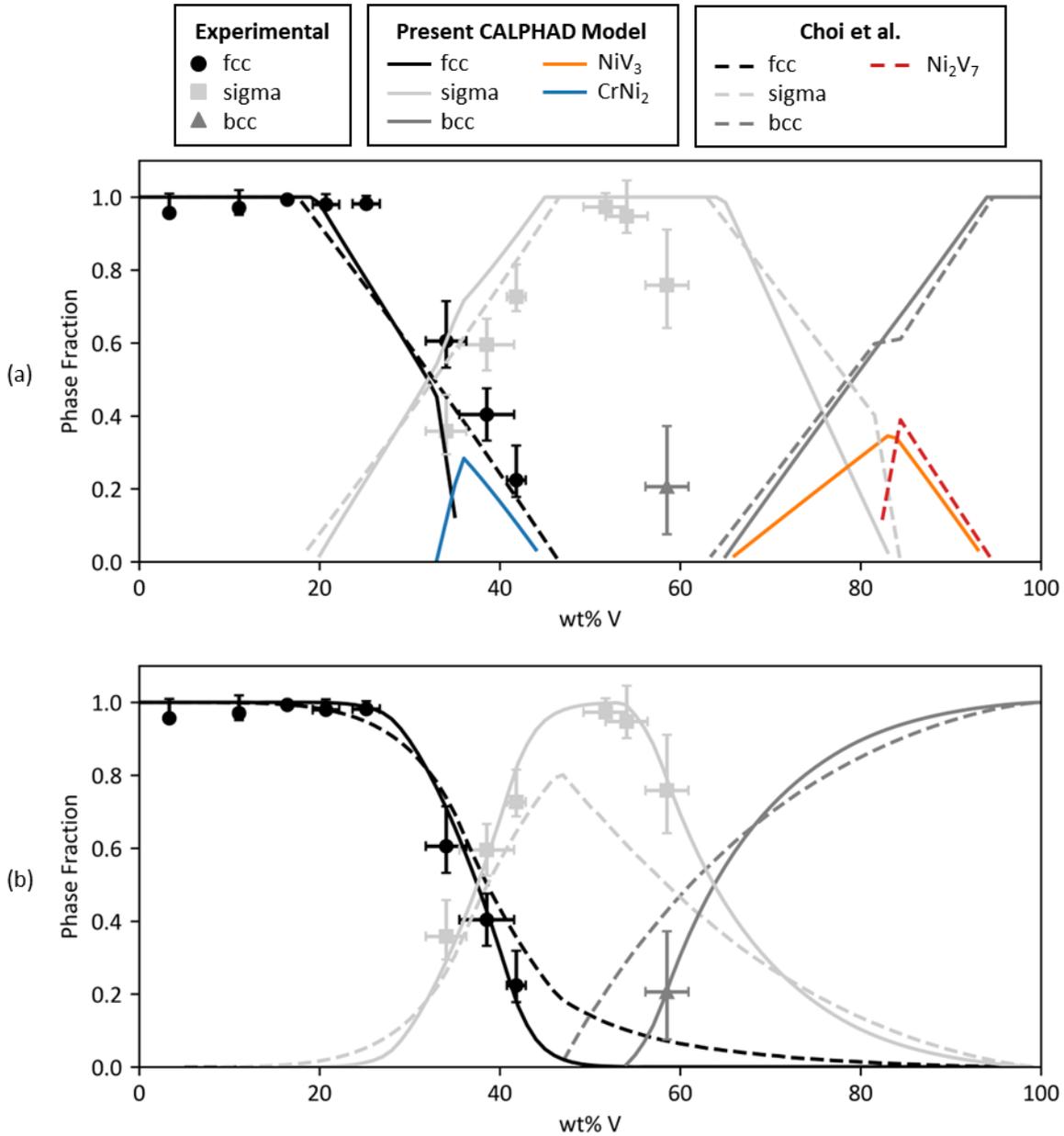

**Figure 1.** Experimentally measured phase mole fractions as a function of wt% V (balance NiCr) of the matrix area in the NiCr-V FGM compared with phases predicted by (a) equilibrium calculations at 1000 K and (b) Scheil solidification simulations using the present database (solid lines) and the database by Choi et al. [13,14] (dashed lines).

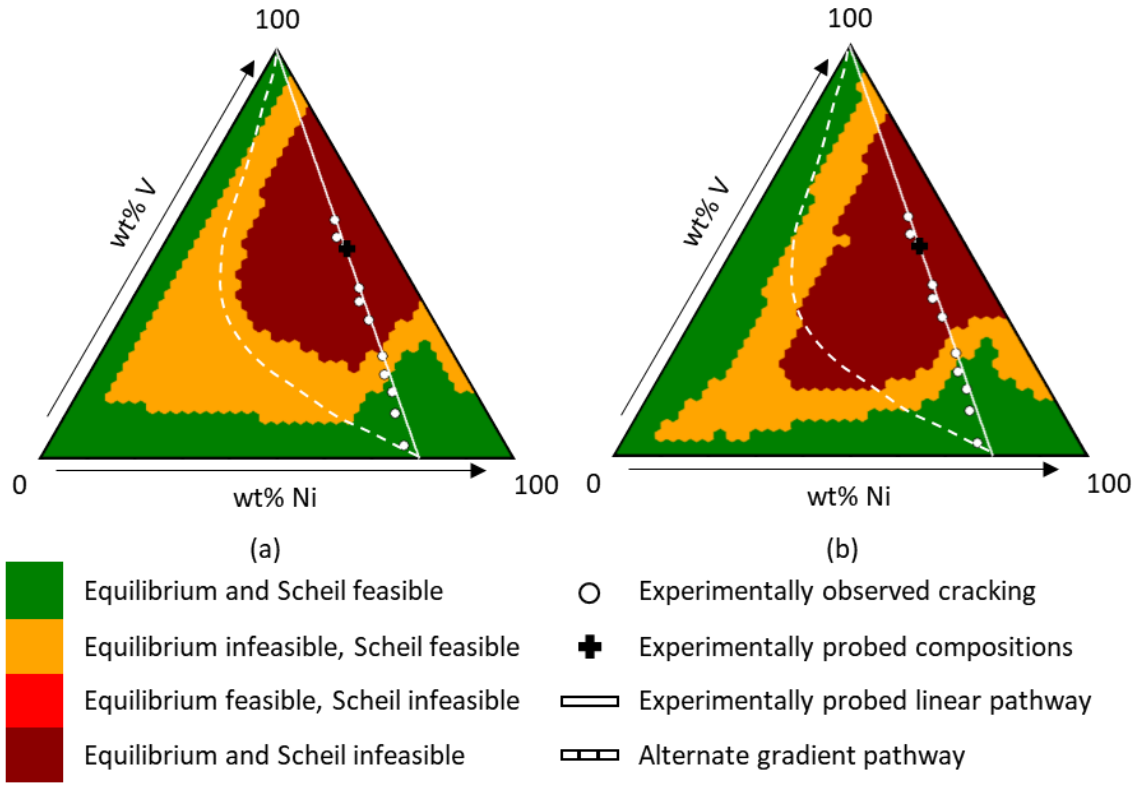

**Figure 2.** Feasibility maps for the Cr-Ni-V ternary showing locations predicted to be feasible based on equilibrium calculations between 1000 K and 2200 K and Scheil calculations using (a) the present database and (b) the database developed in refs. [13,14]. Compositions are assumed to be feasible if under 0.10 mole fraction of Sigma phase was predicted to from under equilibrium and Scheil conditions (areas in green).

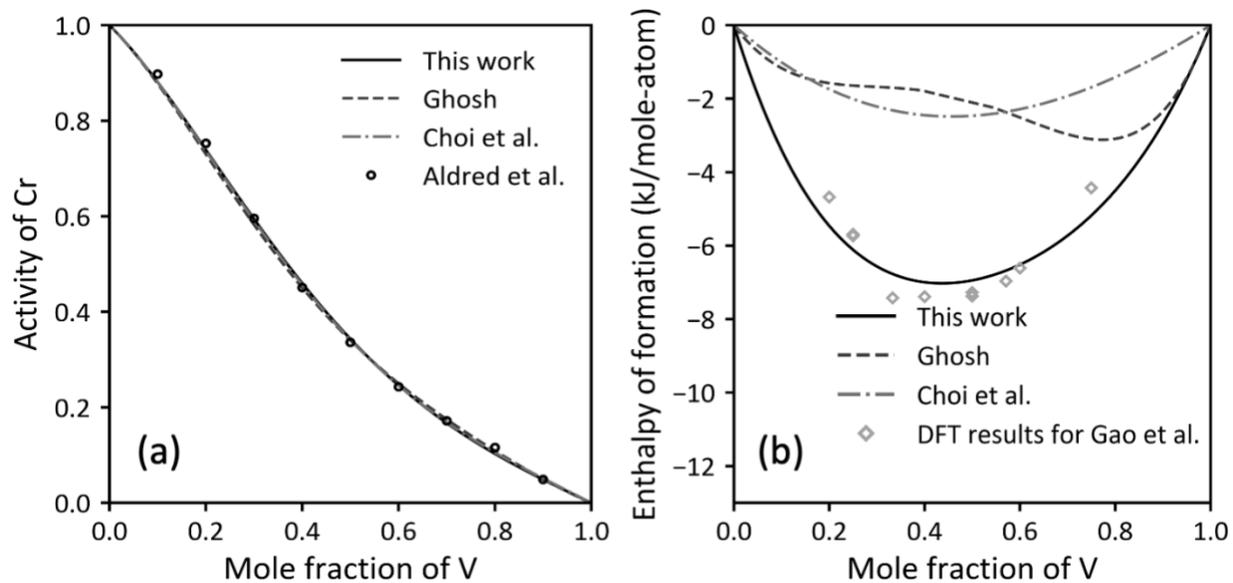

**Figure 3.** Predicted (a) activity of Cr at 1550 K and (b) enthalpy of formation at 298 K using the present database, one developed by Choi et al. [13,14], and a model by Ghosh [32] in comparison with experimental data from (a) Aldred et al. [37] and (b) DFT calculations from Gao et al. [33].

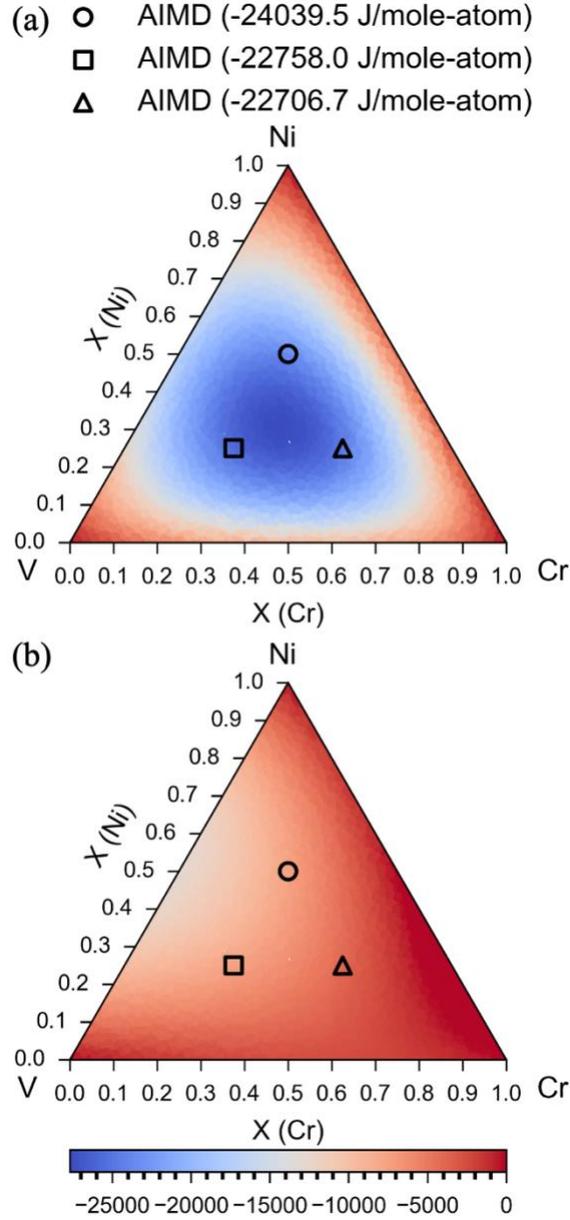

**Figure 4.** Enthalpy of mixing, $\Delta H_{mix}$, of liquid at 2700 K from (a) the present database with differences between CALPHAD and AIMD are 1,184.3 J/mol-atom at 25.0Cr-50.0Ni-25.0V, -112.8 J/mol-atom at 50.0Cr-25.0Ni-25.0V, and -1,334.0 J/mol-atom at 25.0Cr-25.0Ni-50.0V and (b) Choi et al.'s database [13,14] in comparison with AIMD results at 2700 K with differences between CALPHAD and AIMD are 16,441.8 J/mol-atom at 25.0Cr-50.0Ni-25.0V, 18,742.4 J/mol-atom at 50.0Cr-25.0Ni-25.0V, and 15,106.5 J/mol-atom at 25.0Cr-25.0Ni-50.0V

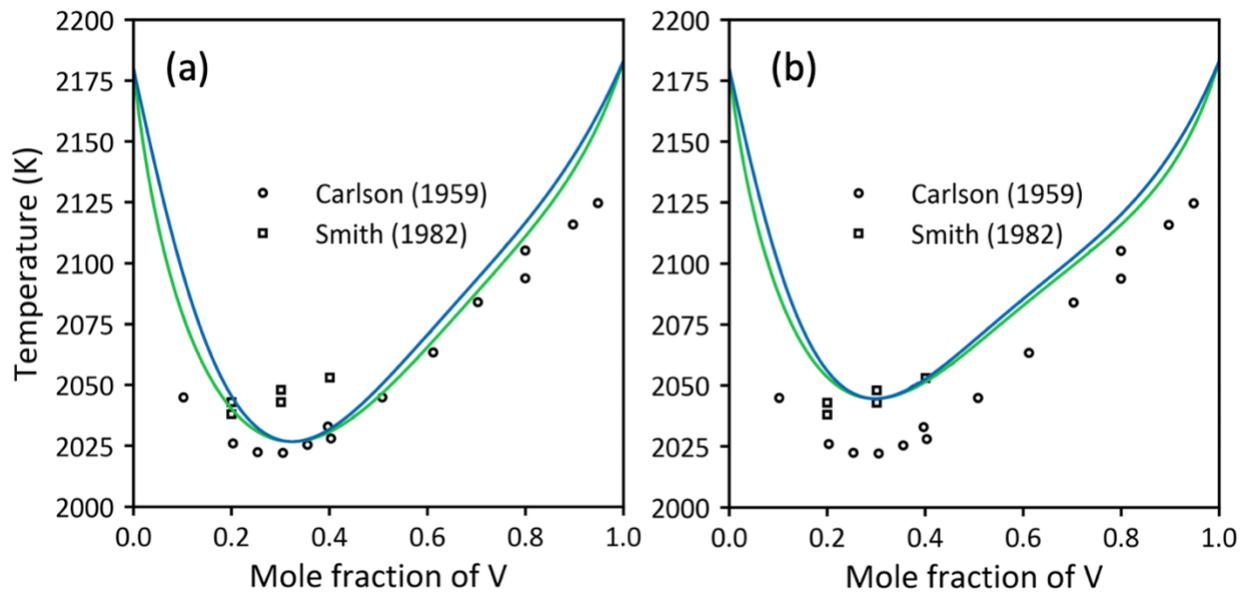

**Figure 5.** Phase diagram of Cr-V using (a) the present database and (b) Ghosh's model [32] with experimental data from Carlson et al. [38] and Smith et al. [40].

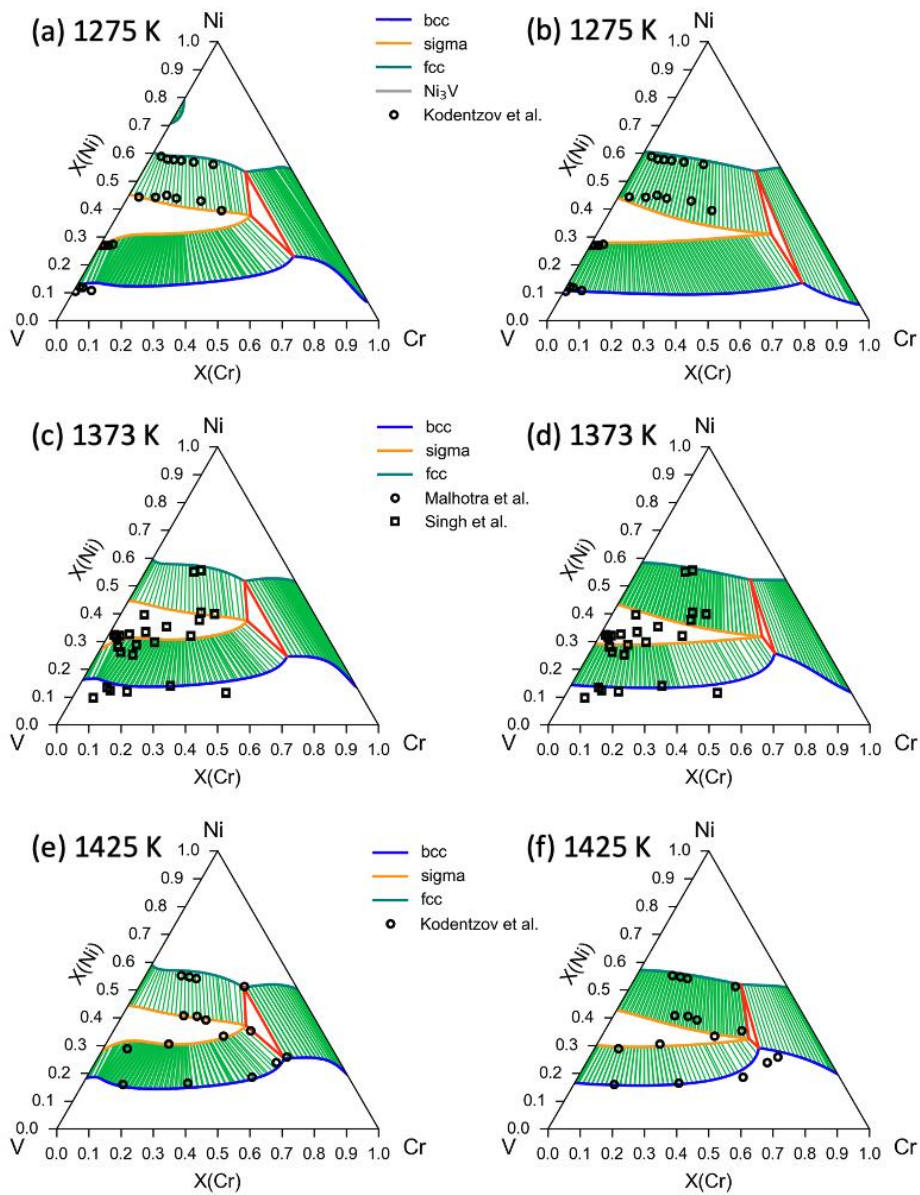

**Figure 6**. Isothermal sections of Cr-Ni-V using both the (a, c, e) present database and (b, d, f) Choi et al. database [13,14] at (a, b) 1275 K, (c, d) 1373 K and (e, f) 1425 K in comparison with experimental data from Kodentzov et al. [48] at 1275 and 1425 K, Singh et al. [49] and Malhotra et al. [50] at 1373 K.

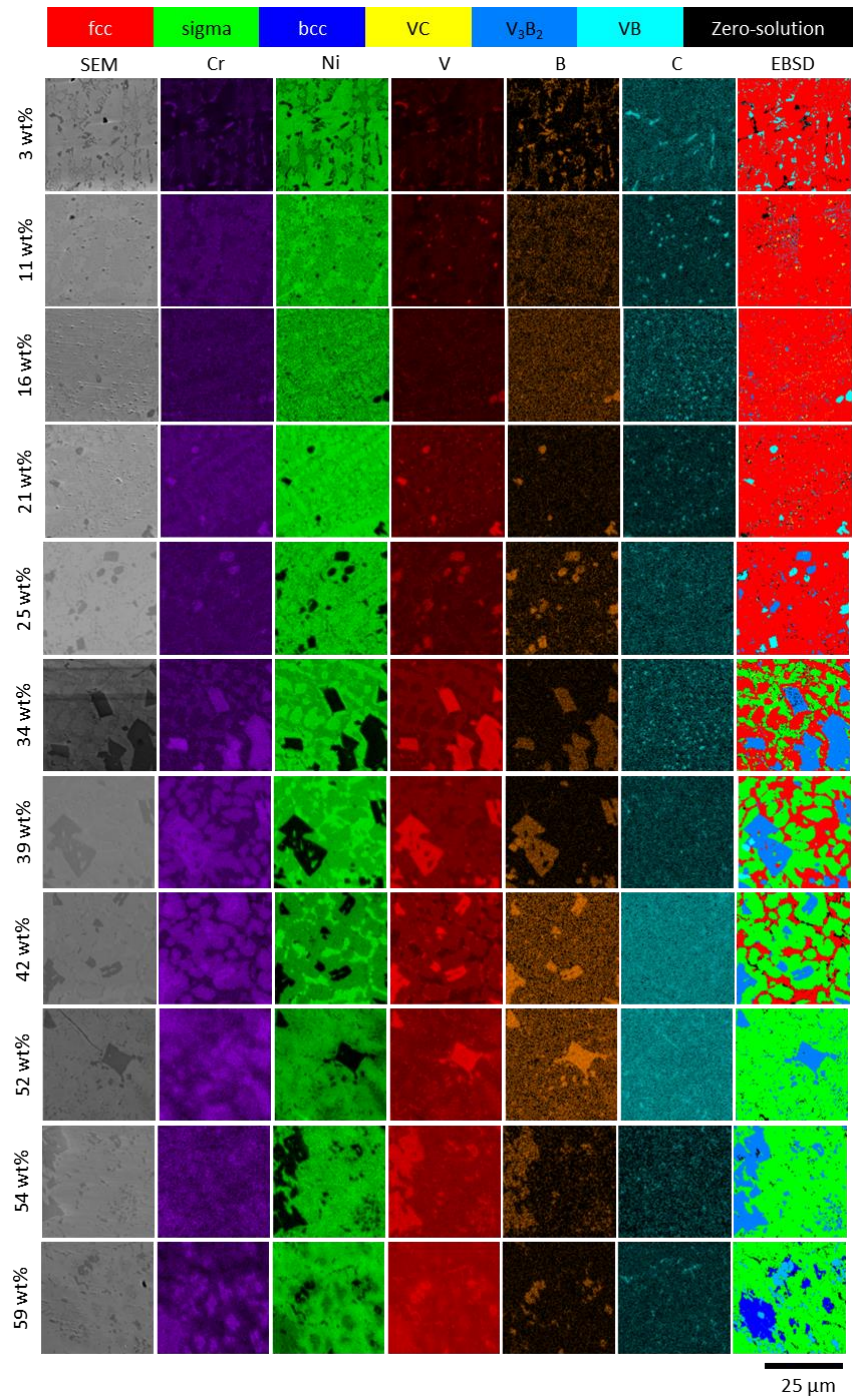

**Figure 7.** Secondary electron images, EDS elemental area maps, and EBSD phase maps of select areas in the NiCr-V FGM sample, where V wt% is given for each row of images. Round V-rich particles are present in regions with 11 – 26 wt% V, while faceted B-, Cr-, and V-rich particles appear with 16-59 wt% V.

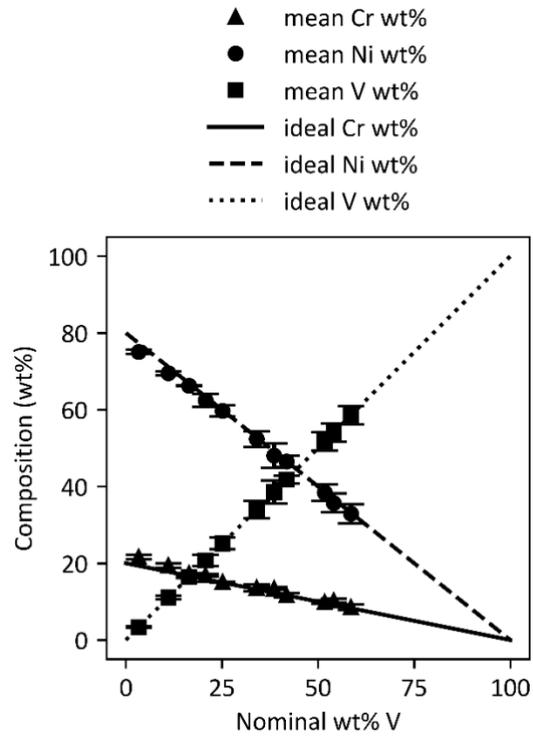

**Figure 8.** Compositions for a nominal linear gradient from NiCr to V (lines) compared to EDS-measured area compositions along the fabricated FGM (symbols), excluding the compositions of the particles.

**Tables**

**Table 1.** Wyckoff positions in coordination x, y, z and their coordination number (CN) of sigma phases.

| Wyckoff position of sigma [a] | x | y | z | CN |
|---|---|---|---|---|
| 2a | 0 | 0 | 0 | 12 |
| 4f | 0.399 | 0.399 | 0 | 15 |
| 8i$_1$ | 0.463 | 0.131 | 0 | 14 |
| 8i$_2$ | 0.739 | 0.066 | 0 | 12 |
| 8j | 0.183 | 0.183 | 0.252 | 14 |

[a] Sigma with space group P4$_2$/mnm (no. 136), Pearson symbol tp30, Strukturbericht designation D8$_b$, and prototype of $\sigma$CrFe [61].

**Table 2.** Settings of DFT-based first-principles calculations, and AIMD calculations for each compound or element.

| Compounds | Atoms in the cells | Cutoff energy (ev) | $k$-points for DFT relaxations | $k$-points for DFT static | $k$-points for AIMD |
|---|---|---|---|---|---|
| Liquid (L) | 108 | 280 | N/A | N/A | 1×1×1 |
| sigma | 30 | 368 (relaxations) 520 (static) | 2×2×3 | 2×2×1 | N/A |

**Table 3.** Processing parameters used to fabricate the Ni-20Cr to V FGM. A laser scan speed of 12.7 mm/s was used throughout.

| Start Layer | End Layer | V vol% | V powder feed rate (g/min) | Ni-20Cr vol% | Ni-20Cr powder feed rate (g/min) | Laser spot size (μm) | Hatch spacing (μm) | Laser power (W) |
|---|---|---|---|---|---|---|---|---|
| 0 | 19 | 0 | 0 | 100 | 5.19 | 1397 | 762 | 800 |
| 20 | 39 | 10 | 0.35 | 90 | 4.67 | 1461 | 826 | 860 |
| 40 | 59 | 20 | 0.70 | 80 | 4.15 | 1524 | 889 | 920 |
| 60 | 79 | 30 | 1.06 | 70 | 3.64 | 1588 | 953 | 980 |
| 80 | 99 | 40 | 1.41 | 60 | 3.12 | 1651 | 1016 | 1040 |
| 100 | 119 | 50 | 1.76 | 50 | 2.60 | 1715 | 1080 | 1100 |
| 120 | 139 | 60 | 2.11 | 40 | 2.08 | 1778 | 1143 | 1160 |
| 140 | 159 | 70 | 2.46 | 30 | 1.56 | 1842 | 1207 | 1220 |
| 160 | 179 | 80 | 2.81 | 20 | 1.04 | 1905 | 1270 | 1280 |
| 180 | 199 | 90 | 3.17 | 10 | 0.52 | 1969 | 1334 | 1340 |
| 200 | 219 | 100 | 3.52 | 0 | 0 | 2032 | 1397 | 1400 |

**Table 4.** Enthalpy of formation values for sigma phases from the present work's DFT-based calculations at 0 K.

| 2a, 8i$_2$ | 4f | 8i$_1$, 8j | ΔH$_f$ (kJ/mol-atom) |
|---|---|---|---|
| Cr | Cr | Cr | 13.670 |
| Cr | Cr | Ni | 6.437 |
| Cr | Cr | V | 5.085 |
| Cr | Ni | Cr | 19.448 |
| Cr | Ni | Ni | 10.516 |
| Cr | Ni | V | 9.394 |
| Cr | V | Cr | 7.876 |
| Cr | V | Ni | 4.503 |
| Cr | V | V | 0.890 |
| Ni | Cr | Cr | 4.072 |
| Ni | Cr | Ni | 10.175 |
| Ni | Cr | V | -9.811 |
| Ni | Ni | Cr | 11.912 |
| Ni | Ni | Ni | 9.843 |
| Ni | Ni | V | -10.582 |
| Ni | V | Cr | -0.347 |
| Ni | V | Ni | -3.164 |
| Ni | V | V | -16.217 |
| V | Cr | Cr | 13.863 |
| V | Cr | Ni | -1.635 |
| V | Cr | V | 6.775 |
| V | Ni | Cr | 19.817 |
| V | Ni | Ni | -5.147 |
| V | Ni | V | 8.540 |
| V | V | Cr | 9.299 |
| V | V | Ni | -7.469 |
| V | V | V | 3.686 |

**Table 5**. Comparison between the experimentally observed phase fractions from EBSD at the given overall area compositions to the computationally predicted area fractions using the B-Cr-Ni-V database.

| Overall Composition (wt%) | | | | Overall Phase Fraction (Experimental) | | | | | | Predicted Phase Fraction | | | | |
|---|---|---|---|---|---|---|---|---|---|---|---|---|---|---|
| B | Cr | Ni | V | zero | fcc | sigma | bcc | m3b2 | mb | fcc | sigma | bcc | m3b2 | mb |
| 1.44 | 25.74 | 68.82 | 4.01 | 0.04 ± 0.01 | 0.88 ± 0.01 | | | | 0.08 ± 0.01 | 0.86 | | | | 0.14 |
| 0 | 18.82 | 67.36 | 10.72 | 0.03 ± 0.02 | 0.94 ± 0.02 | | | 0.01 ± 0.01 | 0.02 ± 0.01 | 1 | | | | |
| 0.43 | 18.09 | 64.21 | 17.27 | 0.01 ± 0 | 0.01 ± 0.01 | | | 0 ± 0.01 | 0.03 ± 0.01 | 0.95 | | | 0.04 | 0.01 |
| 1.26 | 17.69 | 56.43 | 24.61 | 0.02 ± 0.01 | 0.89 ± 0.07 | | | 0.05 ± 0.05 | 0.05 ± 0.06 | 0.85 | | | 0.15 | |
| 1.57 | 15.9 | 53.63 | 28.9 | 0.02 ± 0 | 0.88 ± 0.04 | | | 0.04 ± 0.02 | 0.06 ± 0.05 | 0.8 | 0.01 | | 0.19 | |
| 2.73 | 15.23 | 40.79 | 41.25 | 0.03 ± 0.01 | 0.47 ± 0.14 | 0.27 ± 0.06 | | 0.21 ± 0.18 | 0.01 ± 0.01 | 0.43 | 0.26 | | 0.28 | |
| 1.92 | 13.81 | 38.96 | 41.97 | | 0.34 ± 0.09 | 0.5 ± 0.05 | | 0.14 ± 0.09 | 0.02 ± 0.01 | 0.14 | 0.64 | | 0.22 | |
| 1.43 | 12.23 | 41.13 | 45.21 | 0.04 ± 0.01 | 0.2 ± 0.05 | 0.64 ± 0.03 | | 0.11 ± 0.03 | | 0.15 | 0.68 | | 0.17 | |
| 1.26 | 10.2 | 34.51 | 54.03 | 0.02 ± 0.01 | 0 ± 0 | 0.88 ± 0.05 | | 0.09 ± 0.05 | | | 0.85 | | 0.15 | |
| 2.32 | 10.3 | 29 | 58.37 | 0.05 ± 0.05 | 0 ± 0 | 0.78 ± 0.22 | | 0.18 ± 0.24 | | | 0.68 | 0.06 | 0.26 | |
| 1.29 | 8.63 | 29.51 | 60.57 | 0.03 ± 0.02 | 0 ± 0 | 0.68 ± 0.12 | 0.18 ± 0.11 | 0.11 ± 0.04 | 0 ± 0 | | 0.61 | 0.24 | 0.15 | |

**Table 6.** Calculated overall composition of scan areas, in wt%, along the NiCr-V FGM and their experimentally observed relative ratios of fcc, sigma, and bcc phases compared to the composition of the matrix area, in wt%, with the average phase fraction of the area surrounding the boride particles, referred to as the matrix phase(s).

| Overall Composition (wt%) | | | | Relative Ratio of Average Phase Amounts from Overall Phase Fractions | | | Composition of Matrix (wt%) | | | | Phase Fractions in Matrix | | | |
|---|---|---|---|---|---|---|---|---|---|---|---|---|---|---|
| B | Cr | Ni | V | fcc | sigma | bcc | B | Cr | Ni | V | zero | fcc | sigma | bcc |
| 1.44 | 25.74 | 68.82 | 4.01 | 1 | | | - | 21.6 | 75.07 | 3.33 | 0.04 ± 0.01 | 0.96 ± 0.01 | | |
| 0 | 18.82 | 67.36 | 10.72 | 1 | | | - | 19.42 | 69.51 | 11.07 | 0.03 ± 0.02 | 0.97 ± 0.02 | | |
| 0.43 | 18.09 | 64.21 | 17.27 | 1 | | | - | 17.34 | 66.24 | 16.41 | 0.01 ± 0 | 0.99 ± 0 | | |
| 1.26 | 17.69 | 56.43 | 24.61 | 1 | | | - | 16.87 | 62.42 | 20.71 | 0.02 ± 0.01 | 0.98 ± 0.01 | | |
| 1.57 | 15.9 | 53.63 | 28.9 | 1 | | | - | 15.11 | 59.72 | 25.17 | 0.02 ± 0 | 0.98 ± 0 | | |
| 2.73 | 15.23 | 40.79 | 41.25 | 0.63 | 0.37 | | - | 13.6 | 52.36 | 34.03 | 0.04 ± 0.01 | 0.61 ± 0.07 | 0.36 ± 0.06 | |
| 1.92 | 13.81 | 38.96 | 41.97 | 0.41 | 0.59 | | - | 13.42 | 48.03 | 38.54 | 0 ± 0 | 0.4 ± 0.07 | 0.6 ± 0.07 | |
| 1.43 | 12.23 | 41.13 | 45.21 | 0.24 | 0.76 | | - | 11.67 | 46.51 | 41.82 | 0.05 ± 0.01 | 0.22 ± 0.05 | 0.73 ± 0.04 | |
| 1.26 | 10.2 | 34.51 | 54.03 | 0 | 1 | | - | 9.86 | 38.43 | 51.71 | 0.03 ± 0.01 | | 0.97 ± 0.01 | |
| 2.32 | 10.3 | 29 | 58.37 | 0 | 1 | | - | 10.25 | 35.7 | 54.05 | 0.05 ± 0.05 | | 0.95 ± 0.05 | |
| 1.29 | 8.63 | 29.51 | 60.57 | 0 | 0.79 | 0.21 | - | 8.53 | 32.94 | 58.52 | 0.03 ± 0.02 | | 0.76 ± 0.12 | 0.21 ± 0.13 |

# References

[1]   D.C. Hofmann, J. Kolodziejska, S. Roberts, R. Otis, R.P. Dillon, J.O. Suh, Z.K. Liu, J.P. Borgonia, Compositionally graded metals: A new frontier of additive manufacturing, J Mater Res. 29 (2014) 1899–1910. https://doi.org/10.1557/jmr.2014.208.

[2]   Y. Li, Z. Feng, L. Hao, L. Huang, C. Xin, Y. Wang, E. Bilotti, K. Essa, H. Zhang, Z. Li, F. Yan, T. Peijs, A Review on Functionally Graded Materials and Structures via Additive Manufacturing: From Multi-Scale Design to Versatile Functional Properties, Adv Mater Technol. 5 (2020). https://doi.org/10.1002/admt.201900981.

[3]   R.M. Mahamood, E.T. Akinlabi, Types of Functionally Graded Materials and Their Areas of Application, in: Topics in Mining, Metallurgy and Materials Engineering, Springer Science and Business Media Deutschland GmbH, 2017: pp. 9–21. https://doi.org/10.1007/978-3-319-53756-6_2.

[4]   D.C. Hofmann, S. Roberts, R. Otis, J. Kolodziejska, R.P. Dillon, J. Suh, A.A. Shapiro, Z.-K. Liu, J.-P. Borgonia, Developing Gradient Metal Alloys through Radial Deposition Additive Manufacturing, Scientific Reports 2014 4:1. 4 (2014) 1–8. https://doi.org/10.1038/srep05357.

[5]   C. Zhang, F. Chen, Z. Huang, M. Jia, G. Chen, Y. Ye, Y. Lin, W. Liu, B. Chen, Q. Shen, L. Zhang, E.J. Lavernia, Additive manufacturing of functionally graded materials: A review, Materials Science and Engineering A. 764 (2019). https://doi.org/10.1016/j.msea.2019.138209.

[6]   W. Li, S. Karnati, C. Kriewall, F. Liou, J. Newkirk, K.M. Brown Taminger, W.J. Seufzer, Fabrication and characterization of a functionally graded material from Ti-6Al-4V to SS316 by laser metal deposition, Addit Manuf. 14 (2017) 95–104. https://doi.org/10.1016/j.addma.2016.12.006.

[7]   H.C. Dey, M. Ashfaq, A.K. Bhaduri, K.P. Rao, Joining of titanium to 304L stainless steel by friction welding, J Mater Process Technol. 209 (2009) 5862–5870. https://doi.org/10.1016/j.jmatprotec.2009.06.018.

[8]   H. Sahasrabudhe, R. Harrison, C. Carpenter, A. Bandyopadhyay, Stainless steel to titanium bimetallic structure using LENS$^{TM}$, Addit Manuf. 5 (2015) 1–8. https://doi.org/10.1016/j.addma.2014.10.002.

[9]   L.D. Bobbio, R.A. Otis, J.P. Borgonia, R.P. Dillon, A.A. Shapiro, Z.K. Liu, A.M. Beese, Additive manufacturing of a functionally graded material from Ti-6Al-4V to Invar: Experimental characterization and thermodynamic calculations, Acta Mater. 127 (2017) 133–142. https://doi.org/10.1016/j.actamat.2016.12.070.

[10]  A. Reichardt, R.P. Dillon, J.P. Borgonia, A.A. Shapiro, B.W. McEnerney, T. Momose, P. Hosemann, Development and characterization of Ti-6Al-4V to 304L stainless steel gradient

components fabricated with laser deposition additive manufacturing, Mater Des. 104 (2016) 404–413. https://doi.org/10.1016/j.matdes.2016.05.016.

[11] D. Svetlizky, M. Das, B. Zheng, A.L. Vyatskikh, S. Bose, A. Bandyopadhyay, J.M. Schoenung, E.J. Lavernia, N. Eliaz, Directed energy deposition (DED) additive manufacturing: Physical characteristics, defects, challenges and applications, Materials Today. 49 (2021) 271–295. https://doi.org/10.1016/j.mattod.2021.03.020.

[12] L.D. Bobbio, B. Bocklund, E. Simsek, R.T. Ott, M.J. Kramer, Z.-K. Liu, A.M. Beese, Design of an additively manufactured functionally graded material of 316 stainless steel and Ti-6Al-4V with Ni-20Cr, Cr, and V intermediate compositions, Addit Manuf. 51 (2022) 102649. https://doi.org/10.1016/J.ADDMA.2022.102649.

[13] W.M. Choi, Y.H. Jo, D.G. Kim, S.S. Sohn, S. Lee, B.J. Lee, A Thermodynamic Modelling of the Stability of Sigma Phase in the Cr-Fe-Ni-V High-Entropy Alloy System, J Phase Equilibria Diffus. 39 (2018) 694–701. https://doi.org/10.1007/S11669-018-0672-X/FIGURES/5.

[14] W.M. Choi, Y.H. Jo, D.G. Kim, S.S. Sohn, S. Lee, B.J. Lee, A thermodynamic description of the Co-Cr-Fe-Ni-V system for high-entropy alloy design, Calphad. 66 (2019) 101624. https://doi.org/10.1016/J.CALPHAD.2019.05.001.

[15] Z. Yang, H. Sun, Z.-K. Liu, A.M. Beese, Design methodology for functionally graded materials: Framework for considering cracking, Addit Manuf. 73 (2023) 103672. https://doi.org/10.1016/j.addma.2023.103672.

[16] Thermo-Calc Software TCFE12 Steels/Fe-alloys Database, (n.d.). https://thermocalc.com/products/databases/steel-and-fe-alloys/.

[17] Thermo-Calc Software TCNI12 Nickel-based Superalloys Databases, (n.d.). https://thermocalc.com/support/how-to-cite-thermo-calc-products/.

[18] Thermo-Calc Software TCHEA6 High Entropy Alloys Database, (n.d.). https://thermocalc.com/products/databases/high-entropy-alloys/.

[19] B.-J. Lee, On the stability of Cr carbides, Calphad. 16 (1992) 121–149. https://doi.org/10.1016/0364-5916(92)90002-F.

[20] B.-J. Lee, A Thermodynamic Evaluation of the Fe-Cr-V System/Eine thermodynamische Bewertung des Systems Fe-Cr-V, International Journal of Materials Research. 83 (1992) 292–299.

[21] I. Ansara, A.T. Dinsdale, M.H. Rand, Al-Mg COST 507 Thermochemical database for light metal alloys, (1998).

[22] M. Noori, B. Hallstedt, Thermodynamic modelling of the Ni-V system, CALPHAD. 65 (2019) 273–281. https://doi.org/10.1016/j.calphad.2019.03.010.

[23] J.-M. Joubert, Crystal chemistry and Calphad modeling of the σ phase, Prog Mater Sci. 53 (2008) 528–583. https://doi.org/10.1016/j.pmatsci.2007.04.001.

[24] A. Watson, F.H. Hayes, Some experiences modelling the sigma phase in the Ni–V system, J Alloys Compd. 320 (2001) 199–206. https://doi.org/10.1016/S0925-8388(00)01472-9.

[25] J.M. Joubert, Crystal chemistry and Calphad modeling of the σ phase, Prog Mater Sci. 53 (2008) 528–583. https://doi.org/10.1016/j.pmatsci.2007.04.001.

[26] E. Scheil, Bemerkungen zur schichtkristallbildung, International Journal of Materials Research. 34 (1942) 70–72.

[27] B. Bocklund, L.D. Bobbio, R.A. Otis, A.M. Beese, Z.K. Liu, Experimental validation of Scheil–Gulliver simulations for gradient path planning in additively manufactured functionally graded materials, Materialia (Oxf). 11 (2020). https://doi.org/10.1016/j.mtla.2020.100689.

[28] A. Xue, X. Lin, L. Wang, J. Wang, W. Huang, Influence of trace boron addition on microstructure, tensile properties and their anisotropy of Ti6Al4V fabricated by laser directed energy deposition, Mater Des. 181 (2019) 107943. https://doi.org/10.1016/j.matdes.2019.107943.

[29] A.T. Dinsdale, SGTE data for pure elements, Calphad. 15 (1991) 317–425. https://doi.org/10.1016/0364-5916(91)90030-N.

[30] F. Tang, B. Hallstedt, Using the PARROT module of Thermo-Calc with the Cr–Ni system as example, CALPHAD. 55 (2016) 260–269. https://doi.org/10.1016/j.calphad.2016.10.003.

[31] L. Hao, A. Ruban, W. Xiong, CALPHAD modeling based on Gibbs energy functions from zero kevin and improved magnetic model: A case study on the Cr–Ni system, CALPHAD. 73 (2021) 102268. https://doi.org/10.1016/j.calphad.2021.102268.

[32] G. Ghosh, Thermodynamic and Kinetic Modeling of the Cr-Ti-V System, 23 (2002) 310–328.

[33] M.C. Gao, Y. Suzuki, H. Schweiger, Ö.N. Doğan, J. Hawk, M. Widom, Phase stability and elastic properties of Cr–V alloys, Journal of Physics: Condensed Matter. 25 (2013) 75402.

[34] O. Redlich, A.T. Kister, Algebraic representation of thermodynamic properties and the classification of solutions, Ind Eng Chem. 40 (1948) 345–348.

[35] M. Hillert, The compound energy formalism, J Alloys Compd. 320 (2001) 161–176.

[36] T.B. Massalski, H. Okamoto, Pr. Subramanian, L. Kacprzak, W.W. Scott, Binary alloy phase diagrams, American society for metals Metals Park, OH, 1986.

[37] A.T. Aldred, K.M. Myles, Thermodynamic Properties of Solid Vanadium-Chromium Alloys, TRANSACTIONS OF THE METALLURGICAL SOCIETY OF AIME. 230 (1964) 736.

[38] O.N. Carlson, D.T. Eash, A.L. Eustice, W.R. Clough, Reactive Metals, (1959).

[39] O. Kubaschewski, H. Unal, High Temp, Pressure. 9 (1977) 361.

[40] J.F. Smith, D.M. Bailey, O.N. Carlson, The Cr− V (chromium-vanadium) system, Journal of Phase Equilibria. 2 (1982) 469–473.

[41] S.-L. Shang, Y. Wang, D. Kim, Z.-K. Liu, First-principles thermodynamics from phonon and Debye model: Application to Ni and Ni3Al, Comput Mater Sci. 47 (2010) 1040–1048.

[42] Y. Wang, Z.K. Liu, L.Q. Chen, Thermodynamic properties of Al, Ni, NiAl, and Ni3Al from first-principles calculations, Acta Mater. 52 (2004) 2665–2671. https://doi.org/10.1016/j.actamat.2004.02.014.

[43] G. Kresse, J. Furthmüller, Efficient iterative schemes for ab initio total-energy calculations using a plane-wave basis set, Phys Rev B. 54 (1996) 11169.

[44] P.E. Blöchl, Projector augmented-wave method, Phys Rev B. 50 (1994) 17953–17979. https://doi.org/10.1103/PhysRevB.50.17953.

[45] J.P. Perdew, K. Burke, M. Ernzerhof, Generalized gradient approximation made simple, Phys Rev Lett. 77 (1996) 3865.

[46] S. Nosé, A unified formulation of the constant temperature molecular dynamics methods, J Chem Phys. 81 (1984) 511–519. https://doi.org/10.1063/1.447334.

[47] Y. Kassir, M. Kupiec, A. Shalom, G. Simchen, Cloning and mapping of CDC40, a Saccharomyces cerevisiae gene with a role in DNA repair, Curr Genet. 9 (1985) 253–257. https://doi.org/10.1007/BF00419952.

[48] A.A. Kodentzov, S.F. Dunaev, E.M. Slusarenko, DETERMINATION OF THE PHASE DIAGRAM OF THE V-Ni-Cr SYSTEM USING DIFFUSION COUPLES AND EQUILIBRATED ALLOYS., Journal of the Less-Common Metals. l35 (1987) 15–24. https://doi.org/10.1016/0022-5088(87)90334-1.

[49] S.K. Singh, K.P. Gupta, The Cr-Ni-V system, Journal of Phase Equilibria. 16 (1995) 129–136. https://doi.org/10.1007/BF02664850.

[50] A.K. Malhotra, K.P. Gupta, Phase Equilibria in Cr-Ni-v System at 1100 °{{C}}, 5 (n.d.) 132–135.

[51] R. Otis, Z.-K. Liu, pycalphad: CALPHAD-based Computational Thermodynamics in Python, J Open Res Softw. 5 (2017) 1. https://doi.org/10.5334/jors.140.

[52] P.E.A. Turchi, L. Kaufman, Z.K. Liu, Modeling of Ni-Cr-Mo based alloys: Part II - Kinetics, CALPHAD. 31 (2007) 237–248. https://doi.org/10.1016/j.calphad.2006.12.006.

[53] Yu.B. Kuz'ma, V.S. Telegus, D.A. Kovalyk, X-ray diffraction investigation of the ternary systems V-Cr-B, Nb-Cr-B, and Mo-Cr-B, Soviet Powder Metallurgy and Metal Ceramics. 8 (1969) 403–410. https://doi.org/10.1007/BF00776617.

[54] M. Tojo, T. Tokunaga, H. Ohtani, M. Hasebe, Thermodynamic analysis of phase equilibria in the Cr-Mo-B ternary system, CALPHAD. 34 (2010) 263–270. https://doi.org/10.1016/j.calphad.2010.04.003.

[55] K. Oikawa, N. Ueshima, Experimental Investigation and Thermodynamic Assessment of Ni–B and Ni–B–C Systems, J Phase Equilibria Diffus. (2022) 1–13.

[56] A.A.A.P. da Silva, N. Chaia, F. Ferreira, G.C. Coelho, J.-M. Fiorani, N. David, M. Vilasi, C.A. Nunes, Thermodynamic modeling of the V-Si-B system, Calphad. 59 (2017) 199–206.

[57] A. Jain, S.P. Ong, G. Hautier, W. Chen, W.D. Richards, S. Dacek, S. Cholia, D. Gunter, D. Skinner, G. Ceder, Commentary: The Materials Project: A materials genome approach to accelerating materials innovation, APL Mater. 1 (2013) 011002.

[58] S. Kirklin, J.E. Saal, B. Meredig, A. Thompson, J.W. Doak, M. Aykol, S. Rühl, C. Wolverton, The Open Quantum Materials Database (OQMD): Assessing the accuracy of DFT formation energies, NPJ Comput Mater. 1 (2015). https://doi.org/10.1038/npjcompumats.2015.10.

[59] A.M. Krajewski, J.W. Siegel, J. Xu, Z.K. Liu, Extensible Structure-Informed Prediction of Formation Energy with improved accuracy and usability employing neural networks, Comput Mater Sci. 208 (2022). https://doi.org/10.1016/j.commatsci.2022.111254.

[60] K.E. Spear, P.K. Liao, J.F. Smith, The B-V (Boron-Vanadium) System, Journal of Phase Equilibria. 8 (1987) 447–454.

[61] H.L. Yakel, Atom distributions in sigma phases. I. Fe and Cr atom distributions in a binary sigma phase equilibrated at 1063, 1013 and 923 K, Acta Crystallographica Section B. 39 (1983) 20–28. https://doi.org/10.1107/S0108768183001974.